\documentclass[11pt]{article}
\usepackage{amsmath}
\usepackage{xspace}
\usepackage{amssymb}
\usepackage{epsfig}

\newtheorem{Thm}{Theorem}
\newtheorem{Lem}[Thm]{Lemma}
\newtheorem{Cor}[Thm]{Corollary}
\newtheorem{Prop}[Thm]{Proposition}
\newtheorem{Claim}{Claim}
\newtheorem{Obs}{Observation}

\newtheorem{Def}{Definition}
\newenvironment{proof}{\noindent {\textbf{Proof }}}{$\Box$ \medskip}

\newcommand\B{\{0,1\}}

\newcommand\pr{\mbox{\bf Pr}}

\newcommand {\ie} {\emph{i.e.}\xspace}
\newcommand {\st} {\emph{s.t.}\xspace}
\newcommand {\vs} {\emph{vs.}\xspace}

\begin{document}
\title{\textbf{(Almost) tight bounds for randomized and quantum Local Search
on hypercubes and grids}\thanks{This research was supported in part
by NSF grants CCR-0310466 and CCF-0426582.}}
\author{Shengyu Zhang\footnote{Computer Science Department, Princeton
University, NJ 08544, USA. Email: szhang@cs.princeton.edu}}
\date{}
\maketitle

\abstract{The Local Search problem, which finds a local minimum of
a black-box function on a given graph, is of both practical and
theoretical importance to many areas in computer science and
natural sciences. In this paper, we show that for the Boolean
hypercube $\B^n$, the randomized query complexity of Local Search
is $\Theta(2^{n/2}n^{1/2})$ and the quantum query complexity is
$\Theta(2^{n/3}n^{1/6})$. We also show that for the constant
dimensional grid $[N^{1/d}]^d$, the randomized query complexity is
$\Theta(N^{1/2})$ for $d \geq 4$ and the quantum query complexity
is $\Theta(N^{1/3})$ for $d \geq 6$. New lower bounds for lower
dimensional grids are also given. These improve the previous
results by Aaronson \cite{Aa04}, and Santha and
Szegedy\cite{SS04a}. Finally we show for $[N^{1/2}]^2$ a new upper
bound of $O(N^{1/4}(\log\log N)^{2})$ on the quantum query
complexity, which implies that Local Search on grids exhibits
different properties at low dimensions.}

\section{Introduction} Many important combinatorial
optimization problems arising in both theory and practice are
\textbf{NP}-hard, which forces people to resort to heuristic
searches in practice. One popular approach is local search, in which
one first defines a \emph{neighborhood structure}, then finds a
solution that is locally optimal with respect to this neighborhood
structure. In the past two decades, the local search approach has
been extensively developed and ``has reinforced its position as a
standard approach in combinatorial optimization" in practice
\cite{AHL+97}. Besides the practical applications, local search also
has many connections to the complexity theory, especially to the
complexity classes \textbf{PLS} \footnote{Polynomial Local Search,
introduced by Johnson, Papadimitriou, and Yannakakis \cite{JPY88}.}
and \textbf{TFNP} \footnote{The family of total function problems,
introduced by Megiddo and Papadimitriou \cite{MP91}.}. For example,
the 2SAT-FLIP problem, an important problem known to be complete in
\textbf{PLS}, is actually the local search problem with the
neighborhood structure being the Boolean hypercube $\B^n$ and the
objective function being the sum of the weights of the clauses that
the truth assignment $x\in \B^n$ satisfies. Local search is also
related to physical systems including folding proteins and to the
quantum adiabatic algorithms \cite{Aa04}. We refer readers to the
papers \cite{Aa04,OPS04,SS04a} for more discussions and the book
\cite{AL97} for a comprehensive introduction.

Precisely, the Local Search problem on an undirected graph $G =
(V,E)$ is defined as follows. Given a function $f:V\rightarrow
\mathbb{N}$, find a vertex $v\in V$ such that $f(v)\leq f(w)$ for
all neighbors $w$ of $v$. A class of \emph{generic algorithms} that
has been widely used is as follows: first set out with an initial
point $v\in V$, then repeatedly search the neighbors to find a point
with a smaller $f$ value until it reaches a locally optimal one.
Though empirically this class of algorithms work very well in most
applications, relatively few theoretical results are known about how
good the generic algorithms are, especially for the randomized (and
quantum) algorithms. This paper investigates the Local Search on
some natural neighborhood structures $G$, and proves the optimality
of the generic algorithms for most $G$'s. For some other $G$, we
give an algorithm better than the generic ones.

Among models for the theoretical studies, the query model has drawn
much attention \cite{Aa04,Al83,AK93,LT93,LTT89,SS04a}. In this
model, $f(v)$ can only be accessed by querying $v$, and the
randomized (and quantum) query complexity, denote by $RLS(G)$ (and
$QLS(G)$) is the minimum number of queries needed by a randomized
(and quantum) algorithm that solves the problem. Previously, for
upper bounds on a general $N$-vertex graph $G$, Aldous \cite{Al83}
proved that $RLS(G) = O(\sqrt{N\delta})$ and Aaronson \cite{Aa04}
proved that $QLS(G) = O(N^{1/3}\delta ^{1/6})$, where $\delta$ is
the maximum degree of $G$. Both algorithms are actually the generic
algorithms mentioned above, with the initial point picked as the one
having the minimum $f$ value over some random samples. For lower
bounds, Aaronson \cite{Aa04} considered two special classes of
graphs: the Boolean hypercube $\B^n$ and the constant dimensional
grid $[N^{1/d}]^d$. He showed that for $\B^n$, $RLS(\B^n) =
\Omega(2^{n/2}/n^2)$ and $QLS(\B^n) = \Omega(2^{n/4}/n)$, and that
for $[N^{1/d}]^d$, $RLS([N^{1/d}]^d) = \Omega(N^{1/2 - 1/d}/\log N)$
and $QLS([N^{1/d}]^d) = \Omega(N^{1/4 - 1/(2d)}/\sqrt{\log N})$. It
has also been shown that $QLS([N^{1/2}]^2) = \Omega(N^{1/8})$ by
Santha and Szegedy \cite{SS04a}. However, the final values of $QLS$
and $RLS$ on both types of special graphs remain an open problem,
explicitly stated in an earlier version of \cite{Aa04} and also
(partially) in \cite{SS04a}.

In this paper, we improve these previous results and show tight
bounds on both $RLS$ and $QLS$ in a unified framework. For the
Boolean hypercube, our lower bounds match the known upper bounds
\cite{Aa04, Al83}. For the constant dimensional grid graphs, our
lower bounds also match the known upper bounds except for a few low
dimensional cases. These imply that the generic algorithms
\cite{Aa04,Al83} are the best for all these neighborhood structures.

\begin{Thm} \label{thm: hypercube}
$RLS(\B^n) = \Omega(2^{n/2}n^{1/2}), \quad QLS(\B^n) =
\Omega(2^{n/3}n^{1/6}).$
\end{Thm}


\begin{Thm} \label{thm: grid lb}
$\\RLS([N^{1/d}]^d) =
\begin{cases}
\Omega(N^{1/2}) & \text{if } d \geq 4 \\
\Omega((N/\log N)^{1/2}) & \text{if } d = 3 \\
\Omega\left(N^{1/3}\right) & \text{if } d = 2 \\
\end{cases}, \quad
QLS([N^{1/d}]^d) =
\begin{cases}
\Omega(N^{1/3}) & \text{if } d \geq 6 \\
\Omega((N/\log N)^{1/3}) & \text{if } d =5 \\
\Omega\left(N^{1/2-1/(d+1)}\right) & \text{if } 2\leq d \leq 4 \\
\end{cases}$
\end{Thm}

The proofs for the quantum lower bounds in both theorems use the
quantum adversary method, which was originally proposed by Ambainis
\cite{Am00}, and later generalized in different ways
\cite{Am03,BSS03,LM04,Zh04}. Recently Spalek and Szegedy made the
picture clear by showing that all these generalizations are
equivalent in power \cite{SS04b}. On the other hand, in proving a
particular problem, some of the methods might be easier to use than
the others. In our case, the technique proposed by Zhang \cite{Zh04}
works pretty well.

Inspired by the quantum adversary method, Aaronson gave a technique
called relational adversary method, to prove lower bounds of
randomized query complexity \cite{Aa04}. Our proofs for the
randomized lower bounds will use this method.

Both the quantum adversary method and the relational adversary
method are frameworks of proving lower bounds, parameterized by
input sets and weight functions of input pairs. Both our proofs
and Aaronson's proofs \cite{Aa04} use random walks in the
corresponding graphs to give the input sets and weight functions.
Besides choosing different random walks and different weight
functions, a key innovation that distinguishes our work from
Aaronson's is that we decompose the graph into two parts, the
tensor product of which is the original graph. We perform the
random walk only in one part, and perform a simple one-way walk in
a self-avoiding path in the other part, which serves as a ``clock"
to record the number of steps taken by the random walk in the
first part. The tensor product of these two walks is a random path
in the original graph. A big advantage of adding a clock is that
the ``passing probability", the probability that the random path
\emph{passes} a vertex $v$ \emph{within} $T$ steps, is now the
``stopping probability", the probability that the random walk in
the first part \emph{stops} at $v$ \emph{after} exactly $t$ steps,
which is well understood in the classical random walk literature.
Another advantage is that since the walk in the second part is on
a self-avoiding path, the resulting random path in the original
graph does not intersect with itself either, which makes our
analysis easier.

Finally, we give a new upper bound for $QLS([N^{1/2}]^2)$ by
showing an algorithm working better than the generic algorithms.
Together with the lower bounds in Theorem \ref{thm: grid lb}, this
implies that Local Search on grids exhibits different properties
at low dimensions.

\begin{Thm}\label{thm: grid ub}
$QLS([N^{1/2}]^2) = O(N^{1/4}(\log \log N)^{2})$
\end{Thm}

Both our lower and upper bound techniques can be used on more
general graphs. The proof of Theorem \ref{thm: hypercube} and
\ref{thm: grid lb} generalizes easily to show lower bounds for
Local Search on product graphs. The technique used in the Theorem
\ref{thm: grid ub} can be naturally used on the general graph $G$
that ``expands slowly". See Section \ref{sec: conclusion} for more
detailed discussions.

\vspace{.5em}\noindent\emph{Other related results}. There were two
unpublished results about $RLS([N^{1/2}]^2)$ and $QLS(\B^n)$. It
is mentioned in \cite{Aa04} that Ambainis showed $QLS(\B^n) =
\Omega(2^{n/3}/n^{O(1)})$, and it is mentioned in \cite{SS04a}
that Verhoeven showed $RLS([N^{1/2}]^2) = \Omega(N^{1/2-\delta})$
for any constant $\delta>0$.

\section{Preliminaries and notations}
We use $[M]$ to denote the set $\{1, 2, ..., M\}$. We define the
sign function to be $sign(z) = 1$ if $z> 0$, $-1$ if $z< 0$ and 0
if $z=0$. For an $n$-bit binary string $x = x_0...x_{n-1}\in
\B^n$, let $x^{(i)} = x_0...x_{i-1}(1-x_i)x_{i+1}...x_{n-1}$ be
the string obtained by flipping the coordinate $i$.

A path $X$ in a graph $G = (V,E)$ is a sequence $(v_1, ..., v_l)$
of vertices such that for any pair $(v_i,v_{i+1})$ of vertices,
either $v_i = v_{i+1}$ or $(v_i,v_{i+1})\in E$. We use $set(X)$ to
denote the set of distinct vertices on path $X$.

The $(k,l)$-hypercube $G_{k,l}$ is a special graph whose vertex
set is $V=[k]^l$ and whose edge set is $E=\{(u,v): \exists i\in
[l], \ \st\ |u_i - v_i| = 1, \text{ and }u_j = v_j, \ \forall
j\neq i\}$. Sometimes we abuse the notation by using $[k]^l$ to
denote $G_{k,l}$. Note that both the Boolean hypercube and the
constant dimension grid are special hypercubes.\footnote{Here we
identify the Boolean hypercube $\B^n$ and $G_{2,n}$ since they are
isomorphic.}

In an $N$-vertex graph $G = (V,E)$, a Hamilton path is a path $X =
(v_1, ..., v_{|V|})$ such that $(v_i, v_{i+1})\in E$ for any $i\in
[N-1]$ and $set(X) = V$. It is easy to check by induction that
every hypercube $[k]^l$ has a Hamilton path. Actually, for $l=1$,
$[k]$ has a Hamilton path $(1,...,k)$. Now suppose $[k]^l$ has a
Hamilton path $P$, then a Hamilton path for $[k]^{l+1}$ can be
constructed as follows, first fix the last coordinate to be 1 and
go through $P$, then change the last coordinate to be 2 and go
through $P$ in the reverse order, and change the last coordinate
to be 3 and go through $P$, and so on. For each $(k,l)$, let
$HamPath_{k,l}=(v_1, ..., v_{N})$ be the Hamilton path constructed
as above, and we define the successor function $H_{k,l}(v_i) =
v_{i+1}$ for $i\in [N-1]$.

We use $R_2(f)$ and $Q_2(f)$ to denote the double-sided error random
and quantum query complexities of function $f$. For more details on
query models and query complexities, we refer to \cite{BW02} as an
excellent survey.

\subsection{One quantum adversary method and the relational adversary method}
We describe the quantum adversary method proposed in \cite{Zh04}.
The definition and theorem given here are a little more general
than the original ones, but the proof remains unchanged.

\begin{Def}\label{weight scheme} Let $F: I^N\rightarrow [M]$ be an $N$-variate
function. Let $R\subseteq I^N\times I^N$ be a relation such that
$F(x) \neq F(y)$ for any $(x,y)\in R$. A weight scheme consists of
three weight functions $w(x,y)>0$, $u(x,y,i)>0$ and $v(x,y,i)>0$
satisfying $u(x,y,i)v(x,y,i) \geq w^2(x,y)$ for all $(x,y)\in R$ and
$i\in [N]$ with $x_i\neq y_i$. We further put
\begin{alignat}{2}
w_x & = \sum_{y': (x,y')\in R} w(x,y'), & \qquad w_y & = \sum_{x':
(x',y)\in R} w(x',y) \\
u_{x,i} & =  \sum_{y':(x,y')\in R, x_i \neq y'_i} u(x,y',i), &
\qquad v_{y,i} &= \sum_{x':(x',y)\in R, x'_i \neq y_i} v(x',y,i).
\end{alignat}
\end{Def}

\begin{Thm} \label{thm: Alb4} \emph{[Zhang, \cite{Zh04}]} For any
$F, R$ and any weight scheme $w,u,v$ as in Definition
\ref{weight scheme}, we have
\begin{equation}\label{eq: Alb4}
Q_2(F) =\Omega(\min_{(x,y)\in R,i\in [N], x_i\neq y_i}
\sqrt{\frac{w_xw_y}{u_{x,i}v_{y,i}}})
\end{equation}
\end{Thm}

In \cite{Aa04}, Aaronson gives a nice technique to get a lower bound
for randomized query complexity. We restate it using a similar
language of Theorem \ref{thm: Alb4}.

\begin{Thm} \label{thm: randomized lb} \emph{[Aaronson, \cite{Aa04}]}
Let $F: I^N\rightarrow [M]$ be an $N$-variate function. Let
$R\subseteq I^N\times I^N$ be a relation such that $F(x) \neq F(y)$
for any $(x,y)\in R$. For any weight function $w:R\rightarrow
\mathbb{R}^+$, we have
\begin{equation}\label{eq: relational adversary method}
R_2(F) =\Omega(\min_{(x,y)\in R,i\in [N], x_i\neq y_i}
\max\{\frac{w_x}{w_{x,i}},\frac{w_y}{w_{y,i}}\})
\end{equation}
where
\begin{equation}
w_{x,i} =  \sum_{y':(x,y')\in R, x_i \neq y'_i} w(x,y'), \qquad
w_{y,i} = \sum_{x':(x',y)\in R, x'_i \neq y_i} w(x',y).
\end{equation}
\end{Thm}

Note that we can think of Theorem \ref{thm: randomized lb} as also
having a weight scheme, which requires that $u(x,y,i) = v(x,y,i) =
w(x,y)$. This simple observation will be used in the proof of
Theorem \ref{thm: hypercube} and \ref{thm: grid lb}.

\section{Lower bounds for Local Search on the Boolean Hypercube}
The proof of Theorem \ref{thm: hypercube} uses the following
lemma. Consider that we put $t$ balls randomly into $m$ bins one
by one. The $j$-th ball goes into the $i_j$-th bin. Denote by
$n_i$ the total number of balls in the $i$-th bin. We write
$n_i\equiv b_i$ if $b_i = n_i \text{ mod } 2$. We say that
$(i_1,...,i_t)$ \emph{generates the parity sequence}
$(b_1,...,b_m)$, or simply $(i_1,...,i_t)$ \emph{generates}
$(b_1,...,b_m)$, if $n_i \equiv b_i$ for all $i\in [m]$. For
$b_1...b_m\in\B^m$, denote by $p^{(t)}[b_1,...,b_m]$ the
probability that $n_i \equiv b_i$, $\forall i\in [m]$. We may also
require that the first ball is not put in the bin $i^*$ for some
$i^*\in [m]$. We use $p_{i^*}^{(t)}[b_1,...,b_m]$ to denote the
probability that $n_i \equiv b_i$, $\forall i\in [m]$, under the
condition that the first ball is not put in the bin $i^*$. Let
$p_{i}^{(t)} = \max_{b_1,...,b_m}p_{i}^{(t)}[b_1,...,b_m]$ and
$p_{i^*}^{(t)} = \max_{b_1,...,b_m}p_{i^*}^{(t)}[b_1,...,b_m]$.
The following bounds on $p_{i^*}^{(t)}$ are rather loose but
sufficient for our purpose.
\begin{Lem} \label{lem: hypercube hit prob}
For any $i^*\in [m]$, we have
\begin{equation} p_{i^*}^{(t)} =
\begin{cases}
O(m^{-\lceil t/2 \rceil}) & \text{ if } \ t\leq 10 \\
O(m^{-5}) & \text{ if } \ 10 < t \leq m^2 \\
O(2^{-m}) & \text{ if } \ t > m^2 \\
\end{cases}
\end{equation}
\end{Lem}
The proof of the lemma is in Appendix A. Now we are ready to prove
Theorem \ref{thm: hypercube}.

\vspace{1em}\begin{proof}(of Theorem \ref{thm: hypercube}) We
decompose the whole hypercube $\B^n$ into two spaces $V^w$ and
$V^c$. The first space $V^w$ is an $m$-dimensional hypercube $\B^m$,
where $m$, a fixed value only depending on $d$, will be given later.
The second space $V^c$ is an $(n-m)$-dimensional hypercube
$\B^{n-m}$. Obviously, $\B^n = V^w \otimes V^c$, and each vertex $x
= x_0...x_{n-1}$ in $\B^n$ can be decomposed as $x = x^w\otimes x^c$
where $x^w = x_0...x_{m-1}\in V^w$ and $x^c = x_m...x_{n-1}\in V^c$.
We shall use the two spaces in the following way. In $V^w$ we
perform a random walk; in $V^c$ we set a ``clock", recording how
many steps the random walk in $V^w$ has gone.

Consider the paths $X = (x_{0,0}, x_{0,1}, x_{1,0}, x_{1,1}, ...,
x_{T,0}, x_{T,1})$ where $T=2^{n-m}-1$, that satisfies the
following descriptions.
\begin{enumerate}
\item The starting point $x_{0,0} = x_{0,0}^w\otimes x_{0,0}^c$,
where $x_{0,0}^w$ is any fixed point in $V^w$, say 00...0, and
$x_{0,0}^c$ is the first vertex in the fixed Hamilton path
$HamPath_{2,n-m}$ of $V^c$.

\item For each $t\in \{0,...,T\}$,
\begin{enumerate}
\item $x_{t,1}=(x_{t,0}^w)^{(i_t^x)}\otimes x_{t,0}^c$, where $i_t^x\in \{0,...,
m-1\}$. That is, we randomly choose a coordinate $i_t^x$ of
$x_{t,0}^w$ and flip it.

\item $x_{t+1,0}= x_{t,1}^w\otimes H_{2,n-m}(x_{t,1}^c)$. That is, we let
the clock ``ticks" once.
\end{enumerate}
\end{enumerate}

Let the set $P$ contain all such paths $X$'s, then we define a
problem \textsc{Path$_P$}: given a path $X \in P$, find the end
point $x_{T,1}$. We are allowed to access $X$ by querying an
oracle $O$ whether a point $x\in set(X)$ and getting the Yes/No
answer. Note that an input of \textsc{Path$_P$} is actually a
Boolean function $g: \B^n\rightarrow \B$, with $g(x) = 1$ if and
only if $x\in set(X)$. So strictly speaking, an input should be
specified as $set(X)$ rather than $X$, because in general, it is
possible that $X\neq Y$ but $set(X) = set(Y)$. For our problem,
however, it is easy to check that for any $X,Y\in P$, we have $X =
Y \Leftrightarrow set(X) = set(Y)$. (Actually, if $X\neq Y$,
suppose the first diverging place is $k$, \ie $x_{k,0} = y_{k,0}$,
but $x_{k,1}\neq y_{k,1}$. Then $X$ will never pass $y_{k,1}$
because the clock immediately ticks and the time always advances
forward. Thus $set(X)\neq set(Y)$.) Therefore in what follows, we
shall use $X,Y...$ to specify inputs.

The following claim says that the \textsc{Path$_P$} problem is no
harder than the Local Search.
\begin{Claim}
$R_2(\textsc{Path$_P$})\leq 2 RLS(\B^n)$,
$Q_2(\textsc{Path$_P$})\leq 2 QLS(\B^n)$.
\end{Claim}
\begin{proof}
For any path $X\in P$, we define a function $f_{X}$ essentially in
the same way as Aaronson did in \cite{Aa04}: for each $v\notin X$,
$f_X(v) = \delta(v,x_{0,0}) + 2T$, where $\delta(u,v)$ is the
Hamming distance between $u, v\in \B^n$; for each $x_{t,b}\in
set(X)$, $f_X(x_{t,b}) =2(T-t)-b$. It is easy to check that the
only local minimum point is $x_{T,1}$.

Suppose we have an $Q$-query randomized or quantum algorithm
$\mathcal{A}$ for Local Search, we shall give a $2Q$ algorithm for
\textsc{Path$_P$}. Given an oracle $O$ and an input $X$ of the
\textsc{Path} problem, we run $\mathcal{A}$ to find the local
minimum point of $f_X$, which is also the end point of $X$.
Whenever $\mathcal{A}$ needs to make a query on $v$ to get
$f_X(V)$, it asks $O$ whether $v\in set(X)$. If $v\notin set(X)$,
then $f_X(v) = \delta(v,x_{0,0}) + 2T$; otherwise, $v = x_{t,b}$
for some $t$ and $b$ (note that for a given $x_{t,b}$, $t$ is
fixed and known). If $t=0$, then $f_X(v) = 2T$ if $v=x_{0,0}$ and
$f_X(v) = 2T-1$ otherwise. If $t>0$, then we ask $O$ whether
$v^w\otimes H_{2,n}^{-1}(v^c)\in set(X)$. ($H_{2,n}^{-1}(v)$ gives
the predecessor of $v$ in the fixed Hamilton path.) If yes, then
$v = x_{t,0}$ and thus $f_X(v) = 2(T-t)$; if no, then $v=x_{t,1}$
and $f_X(v) = 2(T-t)-1$. Therefore, at most 2 queries on $O$ can
simulate one query on $f$, so we have a $2Q$ algorithm for
\textsc{Path$_P$}.
\end{proof}

\noindent (Continue the proof of Theorem \ref{thm: hypercube}) By
the claim, it is sufficient to prove lower bounds for
\textsc{Path$_P$}. We define a relation $R_P$ of paths as follows.
\begin{equation}\label{eq: relation}
R_P \hspace{-0.3em}= \{(X,Y)\hspace{-0.3em}:
\hspace{-0.3em}X=(x_{0,0}, x_{0,1}, ..., x_{T,0},
x_{T,1})\hspace{-0.3em}\in P, Y=(y_{0,0}, y_{0,1}, ..., y_{T,0},
y_{T,1})\hspace{-0.3em}\in P, x_{T,1}\neq y_{T,1}\}
\end{equation}

We then choose the weight functions. Recall that for a path $X$,
$i_t^x$ is the coordinate flipped at time $t$. For any $(X,Y)\in
R_P$, we write $X\wedge Y = k$ if $i_0^x = i_0^y$, ..., $i_{k-1}^x
= i_{k-1}^y$ but $i_k^x \neq i_k^y$. Let
\begin{equation}\label{eq: def w}
w(X,Y) = 1/|\{Z\in P: Z\wedge X = k\}|,
\end{equation}
Now let us calculate $w_{X}$. By definition, $w_{X} =
\sum_{Y':(X,Y')\in R_P}w(X,Y')$. We group those $Y'$ that diverge
from $X$ at the same place. Then
\begin{align}
w_{X} & = \sum_{k=0}^{T}\sum_{Y':(X,Y')\in R_P,X\wedge Y' =
k}w(X,Y') \\
& = \sum_{k=0}^{T}\sum_{Y':(X,Y')\in R_P,X\wedge Y' =
k}\frac{1}{|\{Z\in P: Z\wedge X = k\}|} \\
& = \sum_{k=0}^{T}\pr_{Y'\in P}[y'_{T,1}\neq x_{T,1}|Y'\wedge X = k]
\end{align}
By definition, if $Y'\wedge X = T$, then $y'_{T,1}\neq x_{T,1}$ for
sure. If $k<T$, note that for those $Y'$ that $Y'\wedge X = k$,
$y'_{T,1}= x_{T,1}$ if and only if $(i_k^{y'},...,i_T^{y'})$
generates the same parity sequence $(b_1,...,b_m)$ as
$(i_k^x,...,i_T^x)$ does. Thus $\pr_{Y'\in P}[y'_{T,1}\neq
x_{T,1}|Y'\wedge X = k] = 1-p_{i_k^x}^{(T-k+1)}[b_1,...,b_m] =
1-o(1)$ by Lemma \ref{lem: hypercube hit prob}. It follows that
$w_{X} = \sum_{k=0}^{T-1} (1-p_{(i_k^x)}^{(T-k+1)}[b_1,...,b_m]) + 1
= T-o(T)$. Similarly, we have also $w_{Y} = T - o(T)$.

Now we define $u(X,Y,i)$ and $v(X,Y,i)$, where $i$ is a point
$x_{j,b}\in set(X)-set(Y)$ or $y_{j,b} \in set(Y)-set(X)$.
\begin{equation}\label{eq: def u}
u(X,Y,x_{j,b}) = a_{k,j,b}w(X,Y), \qquad u(X,Y,y_{j,b}) =
b_{k,j,b}w(X,Y),
\end{equation}
\begin{equation}\label{eq: def v}
v(X,Y,x_{j,b}) = b_{k,j,b}w(X,Y), \qquad v(X,Y,y_{j,b}) =
a_{k,j,b}w(X,Y).
\end{equation}
where $a_{k,j,b}b_{k,j,b} = 1$, and the values of $a_{k,j,b}$ and
$b_{k,j,b}$ will given later. We now calculate $u_{X,i}$ and
$v_{Y,i}$ for $i = x_{j,b}\in set(X)-set(Y)$ ; the other case
$i=y_{j,b}$ is just symmetric. Note that since $x_{j,b}\in
set(X)-set(Y)$, we have $k\leq j-1$ if $b=0$ and $k\leq j$ if
$b=1$.
\begin{align}
u_{X,x_{j,b}} & = \sum_{k=0}^{j+b-1}\sum_{Y':(X,Y')\in R,\ X\wedge
Y'
= k,\ x_{j,b}\notin set(Y')} a_{k,j,b}w(X,{Y'}) \\
& \leq \sum_{k=0}^{j+b-1}\sum_{Y'\in P:X\wedge Y' = k}
a_{k,j,b}w(X,{Y'}) = \sum_{k=0}^{j+b-1}a_{k,j,b}
\end{align}
The computation for $v_{Y,x_{j,b}}$ is a little more complicated. By
definition,
\begin{align}\label{eq: v(y,i)}
v_{Y,x_{j,b}} & = \sum_{k=0}^{j+b-1}\sum_{X':(X',Y)\in R,\ X'\wedge
Y = k,\ x_{j,b}\in set(X')}
b_{k,j,b}w({X'},{Y}) \\
& \leq \sum_{k=0}^{j+b-1}\sum_{X'\in P:X'\wedge Y = k,\ x_{j,b}\in
set(X')}
b_{k,j,b}w({X'},{Y}) \\
& = \sum_{k=0}^{j+b-1}b_{k,j,b}\pr_{X'\in P}[x_{j,b}\in
set(X')|X'\wedge Y = k]
\end{align}
Note that because of the clock, $x_{j,b}\in set(X')$ if and only if
$x_{j,b} = x'_{j,b'}$ for some $b'\in \B$. And actually $b = b'$,
because otherwise $x_{j,b}$ and $x_{j,b'}$ have different parities
of number of 1's. Therefore, $\pr_{X'}[x_{j,b}\in set(X')|X'\wedge Y
= k] = \pr_{X'}[x_{j,b} = x'_{j,b} |X'\wedge Y = k] =
p_{i_k^y}^{(j-k+b)}[b_1,...,b_m] \leq p_{i_k^y}^{(j-k+b)}$, where
$(b_1,...,b_m)$ is the parity sequence generated by
$i_k^x,...,i_{j+b-1}^x$. So
\begin{align}
v_{Y,x_{j,b}} & \leq \sum_{k=0}^{j+b-1}b_{k,j,b}p_{i_k^y}^{(j-k+b)}
=
O(\sum_{k=0}^{j-m^2+b-1}b_{k,j,b}/2^m+\sum_{k=j-m^2+b}^{j+b-11}b_{k,j,b}/m^5+
\sum_{k=j+b-10}^{j+b-1}b_{k,j,b}/m^{\lceil (j-k+b)/2 \rceil})
\end{align}

Now for the randomized lower bound purpose, we pick $m=\lfloor
(n+\log_2 n)/2 \rfloor$, $a_{k,j,b} = b_{k,j,b} = 1$. Then $T =
2^{n-m} -1 = \Theta(2^{n/2}/\sqrt{n})$, $\frac{w_{X}}{u_{X,x_{j,b}}}
= \frac{T-o(T)}{j} \geq 1-o(1)$, and
\begin{equation}
\frac{w_{Y}}{v_{Y,x_{j,b}}}
=\Omega\left(\frac{T-o(T)}{\frac{j}{2^m}+m^2/m^5 + \sum_{t=1}^5
1/m^t}\right) =
\Omega\left(\frac{2^{n/2}/\sqrt{n}}{\frac{2^{n/2}/\sqrt{n}}{\sqrt{n}2^{n/2}}+1/n}\right)
= \Omega(\sqrt{n}2^{n/2}).
\end{equation}
It is easy to check using the same calculations that for any
$y_{j,b}\in set(Y)-set(X)$, $\frac{w_{X}}{u_{X,y_{j,b}}} =
\Omega(\sqrt{n}2^{n/2})$, and $\frac{w_{Y}}{v_{Y,y_{j,b}}}
\geq 1-o(1)$. Therefore, in either case ($i=x_{j,b}$ or
$i=y_{j,b}$), we have
\begin{equation}
RLS(\B^n) = \max\{\frac{w_{X}}{u_{X,i}}, \frac{w_{Y}}{v_{Y,i}}\} =
\Omega(\sqrt{n}2^{n/2})
\end{equation}

For the quantum lower bound, we pick $m=\lfloor(2n-\log n)/3
\rfloor$, and
\begin{equation}
a_{k,j,b} =
\begin{cases}
m^{-\lceil (j-k+b)/2\rceil/2} & \text{if } j-k+b\leq 10 \\
m^{-5/2} & \text{if } 10< j-k+b\leq m^2 \\
2^{-m/2} & \text{if } j-k+b > m^2 \\
\end{cases}
,\quad b_{k,j,b} =
\begin{cases}
m^{\lceil (j-k+b)/2\rceil/2} & \text{if } j-k+b\leq 10 \\
m^{5/2} & \text{if } 10< j-k+b\leq m^2 \\
2^{m/2} & \text{if } j-k+b > m^2 \\
\end{cases}
\end{equation}
Clearly $a_{k,j,b}b_{k,j,b} = 1$ holds. Note that $T = 2^{n-m}-1 =
\Theta(2^{n/3}n^{2/3})$. Thus $w_{X} = w_{Y} = \Omega(T) =
\Omega(2^{n/3}n^{2/3})$, and
\begin{align}\label{eq:u}
u_{X,x_{j,b}} & \leq \sum_{k=0}^{j+b-1}a_{k,j,b} =
\sum_{k=0}^{j-m^2+b-1} 2^{-m/2} + \sum_{k=j-m^2+b}^{j+b-11}m^{-5/2}
+ \sum_{k=j+b-10}^{j+b-1} m^{-\lceil(j-k+b)/2\rceil /2} =
O(\sqrt{n})
\end{align}
\begin{align}\label{eq:v}
v_{Y,x_{j,b}} & \leq O(\sum_{k=0}^{j-m^2+b-1}2^{m/2}/2^{m} +
\sum_{k=j-m^2+b}^{j+b-11} m^{5/2}/m^5 + \sum_{k=j+b-10}^{j+b-1}
m^{\lceil(j-k-1) /2\rceil /2}/2^{\lceil (j-k-1)/2 \rceil}) =
O(\sqrt{n})
\end{align}
It is easy to check that the above inequalities all hold for the
symmetric case of $y_{j,b}$, so
\begin{equation}
QLS(\B^n) =
\Omega\left(\sqrt{\frac{(2^{n/3}n^{2/3})(2^{n/3}n^{2/3})}{O(\sqrt{n})O(\sqrt{n})}}\right)
= \Omega(2^{n/3}n^{1/6}).
\end{equation}
\end{proof}

\section{Lower bounds for Local Search on the constant dimensional
grid} To simplify notations, we let $n=N^{1/d}$. For $x =
x_0...x_{d-1}$ in $[n]^d$, let $x^{(k)= l} =
x_0...x_{k-1}lx_{k+1}...x_{d-1}$, and $x^{(k) = (k) + i} =
x_0...x_{k-1}(x_k+i)x_{k+1}...x_{d-1}$, where $i$ satisfies
$x_k+i\in [n]$. Also let $x^{(i),-} = x^{(i)=\max\{x_i-1,1\}}$ and
$x^{(i),+} = x^{(i)=\min\{x_i+1,n\}}$.

\subsection{1-dimensional short walk}

We will use random walk on an $n$-point line, where a particle is
initially put at point $i\in \{1, ..., n\}$, and in each step the
particle moves either to $\max\{1, i-1\}$ or to $\min\{n, i+1\}$
with equal probability. That is, the particle randomly choose to
move left or right, but if it is currently at the left (or right)
end and still wants to move left (or right), then it stands still.
We refer to it as short walk. Let $p_{ij}^{(t)}$ denote the
probability that the particle starting from point $i$ stops at
point $j$ after exact $t$ steps of the walk. Obviously, we have
$\max_{i,j}p_{ij}^{(t)} = 1$ if $t=0$. For $t\geq 1$, the
following proposition gives a good estimate on
$\max_{ij}p_{ij}^{(t)}$.
\begin{Prop}\label{prop: short walk mixing}
For any $t\geq 1$,
\begin{equation}
\max_{i,j}p_{ij}^{(t)} =
\begin{cases}
O(1/\sqrt{t}) & \text{if}\quad t\leq n^2 \\
O(1/n) & \text{if}\quad t > n^2 \\
\end{cases}
\end{equation}
\end{Prop}
The proof of the proposition is in Appendix B.

\subsection{Weaker lower bounds}\label{sec: weak grid}
We shall first show a weaker result in this section, then we
improve it in section \ref{sec: better grid}. As in the proof of
Theorem \ref{thm: hypercube}, we decompose the space $[n]^d$ into
two parts $V^w \otimes V^c$, where $V^w = [n]^m$ and $V^c =
[n]^c$. Each vertex $x = x_0...x_{d-1}$ in $[n]^d$ can be
decomposed as $x = x^w\otimes x^c$ where $x^w = x_0...x_{m-1}\in
V^w$ and $x^c = x_{m}...x_{d-1}\in V^c$. Consider the paths $X =
(x_{0,0}, x_{0,1}, x_{1,0}, x_{1,1}, ..., x_{T,0}, x_{T,1})$,
where $T=n^{d-m}-1$, satisfying the following description.
\begin{enumerate}
\item The starting point $x_{0,0} = x_{0,0}^w\otimes x_{0,0}^c$,
where all coordinates of $x_{0,0}^w$ are $\lfloor n/2 \rfloor$, and
$x_{0,0}^c$ is the first vertex in the fixed Hamilton path
$HamPath_{n,d-m}$ of $V^c$.

\item For each $t\in \{0,...,T\}$,
\begin{enumerate}
\item $x_{t,1}\in \{x_{t,0}^{(t \text{ mod } m), +}, x_{t,0}^{(t
\text{ mod } m), -}\}$.

\item $x_{t+1,0}= x_{t,1}^w\otimes H(x_{t,1}^c)$.
\end{enumerate}
\end{enumerate}

Let $P$ contain all such paths $X$'s, then we define the
\textsc{Path$_P$} problem in the same way as in the proof of
Theorem \ref{thm: hypercube}, and it is easy to show that
$R_2(\textsc{Path$_P$}) \leq 2RLS([n]^d)$ and
$Q_2(\textsc{Path$_P$}) \leq 2QLS([n]^d)$. We write $X\wedge Y =
k$ if $x_{0,0} = y_{0,0}$, $x_{0,1} = y_{0,1}$, ..., $x_{k,0} =
y_{k,0}$ but $x_{k,1}\neq y_{k,1}$. We then define $R_P$ and all
weight functions $w, u, v$ in the same form as those in the proof
of Theorem \ref{thm: hypercube} (\ie \eqref{eq:
relation}\eqref{eq: def w}\eqref{eq: def u}\eqref{eq: def v}). For
two points $z_1,z_2\in V^w$, define $z_1\rightarrow_{t}^l z_2$ to
be the event that a random walk starting at $z_1$ stops at $z_2$
after exact $t$ steps, performing one step of short walk in
dimension ($(l+s-1)$ mod $m$) in the $s$-th step $(s\in [t])$. By
Proposition \ref{prop: short walk mixing}, we know that
$\pr[z_1\rightarrow_{t}^l z_2] =
O(\frac{1}{\sqrt{t/m}^m})=O(\frac{1}{\sqrt{t}^m})$ if $1\leq t
\leq mn^2$, and $\pr[z_1\rightarrow_{t}^l z_2] = O(1/n^m)$ if $t
> mn^2$. By some calculations similar to those in the proof of Theorem
\ref{thm: hypercube}, we have $w_X = w_Y = T-o(T)$,
$u_{X,x_{j,b}}\leq \sum_{k=0}^{j+b-1}a_{k,j,b}$, and $v_{Y,x_{j,b}}
\leq \sum_{k=0}^{j+b-1}b_{k,j,b}\pr_{X'}[x_{j,b}\in set(X')|X'\wedge
Y = k]$. Note that $x_{j,b}\in set(X')\Leftrightarrow x_{j,b} =
x'_{j,b}$ again due to the clock and the parity. Also note that if
$X'\wedge Y = k$, then $x_{j,b} = x'_{j,b}\Leftrightarrow
x_{k,1}^w\rightarrow_{j-k+b-1}^{(k+1)\text{ mod } m} x_{j,b}^w$.
Therefore,
\begin{align}
v_{Y,x_{j,b}} & \leq \sum_{k=0}^{j+b-1} b_{k,j,b}\pr[x_{k,1}^w
\rightarrow_{j-k+b-1}^{(k+1)\text{ mod } m} x_{j,b}^w] \\
\label{eq: 3 items} & =
O\left(\sum_{k=0}^{j-mn^2+b-1}\frac{b_{k,j,b}}{n^m} +
\sum_{k=j-mn^2+b}^{j+b-2}\frac{b_{k,j,b}}{\sqrt{(j-k+b-1)^m}}+b_{j+b-1,j,b}\right)
\end{align}
Now for the randomized lower bound purpose, we take $a_{k,j,b} =
b_{k,j,b} = 1$. Then $\frac{w_{X}}{u_{X,x_{j,b}}} = \Omega(1)$, and
$v_{Y,x_{j,b}} = O\left(\frac{T}{n^m} +
\sum_{i=1}^{mn^2}i^{-m/2}+1\right) = O\left(n^{d-2m} +
\sum_{i=1}^{mn^2}i^{-m/2}+1\right)$. When $d>4$ we pick $m=\lceil
d/2 \rceil> 2$, then $v_{Y,x_{j,b}} = O(n^{d-2m})+O(1)$ and
$\frac{w_{Y}}{v_{Y,x_{j,b}}} = \Omega(n^{\lfloor d/2\rfloor})$.
Therefore
\begin{equation}
RLS([n]^d) = \Omega(R_2(\textsc{Path$_P$})) =
\Omega(\max\{\frac{w_{X}}{u_{X,x_{j,b}}}, \frac{w_{Y}}{v_{Y,x_{j,b}}
}\}) =
\begin{cases}
\Omega(N^{\frac{1}{2}}) & \text{ if } d=2d' \\
\Omega(N^{\frac{1}{2}-\frac{1}{2d}}) & \text{ if } d=2d'+1
\end{cases}
\end{equation}
For $d = 4$ and $3$, we let $m=2$ and get $ RLS([n]^4) =
\Omega(n^2/\log n) = \Omega(N^{1/2}/\log N)$ and $ RLS([n]^3) =
\Omega(n) = \Omega(N^{1/3})$. For $d = 2$, we let $m = 1$ and note
that now the walk has only $n$ long, so $w_Y = \Theta(n)$,
$v_{Y,x_{j,b}} = O(\sqrt{n})$, and so $RLS([n]^2) = \Omega(\sqrt{n})
= \Omega(N^{1/4})$.

For the quantum lower bounds, take
\begin{equation}\label{eq: grid ab}
a_{k,j,b} =
\begin{cases}
1 & \text{if } j-k+b = 1 \\
(j-k+b-1)^{-m/4} & \text{if } 1< j-k+b \leq mn^2 \\
n^{-m/2} & \text{if } j-k+b > mn^2
\end{cases}
\quad b_{k,j,b} =
\begin{cases}
1 & \text{if } j-k+b = 1 \\
(j-k+b-1)^{m/4} & \text{if } 1< j-k+b \leq mn^2 \\
n^{m/2} & \text{if } j-k+b > mn^2
\end{cases}
\end{equation}
Then $u_{X,x_{j,b}}= v_{Y,x_{j,b}} =O\left( Tn^{-m/2} +
\sum_{i=1}^{n^2}i^{-m/4}\right)$, and
\begin{equation}\label{eq: grid qlb in m}
QLS([n]^d) = \Omega(Q_2(\textsc{Path$_P$})) =
\Omega(\sqrt{\frac{w_{X}w_{Y}}{u_{X,x_{j,b}}v_{Y,x_{j,b}}}}) =
\Omega\left(\frac{T}{Tn^{-m/2} + \sum_{i=1}^{n^2}i^{-m/4}}\right)
\end{equation}
If $d>6$, then we let $m$ be the integer closest to $2d/3$, thus
$m>4$. We get
\begin{equation}
QLS([n]^d) =
\begin{cases}
\Omega(N^{\frac{1}{3}}) & \text{ if } d = 3d' \\
\Omega(N^{\frac{1}{3}-\frac{1}{3d}}) & \text{ if } d = 3d'+1 \\
\Omega(N^{\frac{1}{3}-\frac{1}{6d}}) & \text{ if } d = 3d'+2 \\
\end{cases}.
\end{equation}
For $d=6$, let $m=4$ and we have $QLS([n]^6) = \Omega(n^2/\log n ) =
\Omega(N^{1/3}/\log N)$. For $d = 5, 4, 3$, we let $m=d-2$ and then
$w_Y = \Theta(n^2)$, $v_{Y,x_{j,b}} = O(n^{3-d/2})$ and $QLS([n]^d)
= \Omega(n^{d/2-1}) = \Omega(N^{1/2-1/d})$. For $d=2$, let $m=1$ and
$QLS([n]^2) = \Omega(\frac{n}{n^{3/4}}) = \Omega(n^{1/4}) =
\Omega(N^{1/8})$.

\subsection{Improvement}\label{sec: better grid} One weakness of
the above proof is the integer constraint of the dimension $m$. We
now show a way to avoid the problem. The idea is to partition the
grid into many blocks, and different blocks represent different
time slots.

\begin{figure}[h]
\begin{center}
\epsfig{file=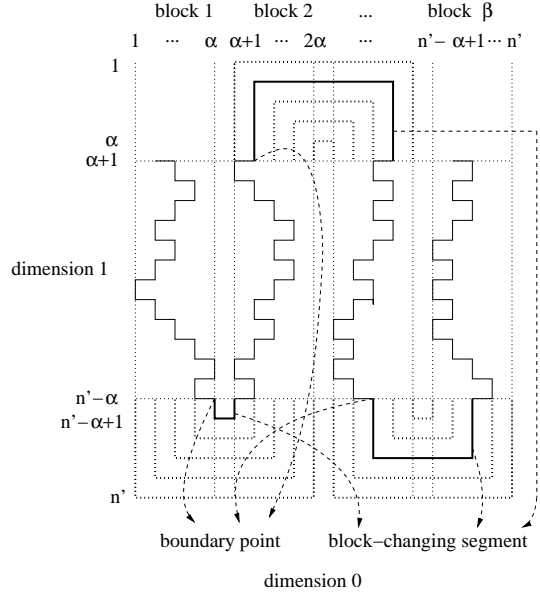, width=7cm} \caption{Illustration for
changing a block in 2 dimensional grid}
\end{center}
\end{figure}

For any fixed $r$, where $r$ will be given later, let $\alpha =
\lfloor n^r \rfloor$, $\beta = \lfloor n^{1-r} \rfloor$ and $n' =
\alpha\beta$. We now consider the slightly smaller grid $[n']^d$.
Let $V_1$ be the set $\{x_0...x_{d-2}: x_i\in [n']\}$. We cut
$V_1$ into $\beta^{d-1}$ parts, each of which is a small grid
$[\alpha]^{d-1}$. We refer to the set $\{x_0...x_{d-2}x_{d-1}:
(k_i-1)\alpha < x_i \leq k_i\alpha, i = 0,...,d-2, \alpha <
x_{d-1}\leq n'-\alpha\}$ as the \emph{block} $(k_0,...,k_{d-2})$.
Note that $(k_0,...,k_{d-2})$ can be also viewed as a point in
grid $[\beta]^{d-1}$, and there is a Hamilton path
$HamPath_{\beta,d-1}$ in $[\beta]^{d-1}$, as defined in Section 2.
We call the block $(k'_0,...,k'_{d-2})$ \emph{the next block} of
the block $(k_0, ..., k_{d-2})$ if $(k'_0,...,k'_{d-2})$, viewed
as the point in $[\beta]^{d-1}$, is the next point of $(k_0, ...,
k_{d-2})$ in $HamPath_{\beta,d-1}$. Note that in
$HamPath_{\beta,d-1}$, to go to the point next to $(k_0, ...,
k_{d-2})$, only one of $k_0, ..., k_{d-2}$ changes by increasing
or decreasing by 1. We call the the block $(k_0, ..., k_{d-2})$
\emph{the last block} if $(k_0, ..., k_{d-2})$ is the last point
in $HamPath_{\beta,d-1}$.

Now we define the random walk by describing how a particle may go
from start to end. The path set is just all the possible paths the
particle goes along. Intuitively, within one block, the last
dimension $d-1$ is the clock space as before. If we run out of it,
we say we reach \emph{a boundary point} at the current block, and
we change to the next block via a path segment called
\emph{block-changing segment}. In what follows, we always use
$x_0...x_{d-1}$ to denote the current position of the particle.
Thus $x_0 = x_0 + 1$, for example, means the particle moves from
$x_0...x_{d-1}$ to $(x_0+1)x_1...x_{d-1}$. We also use
$(k_0,...,k_{d-2})$ to denote the current block which the particle
is in.
\begin{enumerate}
\item Initially  $x_0 = ... = x_{d-2} = \lfloor \alpha/2 \rfloor$,
$x_{d-1} = \alpha+1$, $k_0 = ... = k_{d-2} = 1$.

\item \textbf{for} $t = 0$ \textbf{to}
$(n'-2\alpha)\beta^{d-1}-1$,

\quad Let $t' = \lfloor \frac{t}{n'-2\alpha} \rfloor$, $i=t \text{
mod } (d-1)$

\quad \textbf{do} either $x_{i} = \max\{x_{i} - 1,
(k_i-1)\alpha+1\}$ or $x_{i} = \min\{x_{i} + 1, k_i\alpha\}$
randomly

\quad \textbf{if} $t \neq k(n'-2\alpha)-1$ for some positive
integer $k$,

\quad \quad $x_{d-1} = x_{d-1}+(-1)^{t'}$

\quad \textbf{else}

\quad \quad \textbf{if} the particle is not in the last block
\quad

\quad \quad (Suppose the current block changes to the next block
by increasing $k_j$ by $b\in \{-1, 1\}$)

\quad \quad \quad \textbf{do} $x_{d-1} = x_{d-1}+(-1)^{t'}$
\textbf{for} $(\alpha+1-x_j)$ times

\quad \quad \quad \textbf{do} $x_{j} = x_{j}+b$ \textbf{for}
$2(\alpha+1-x_j)-1$ times

\quad \quad \quad \textbf{do} $x_{d-1} = x_{d-1}+(-1)^{t'+1}$
\textbf{for} $(\alpha+1-x_j)$ times

\quad \quad \quad $k_j = k_j + b$

\quad \quad \textbf{else}

\quad \quad \quad The particle stops and the random walk ends

\end{enumerate}

It is easy to check that every boundary point has one unique
block-changing segment, and different block-changing segments do
not intersect. Thus the block-changing segments thread all the
blocks to form a $[\alpha]^{d-1}\times [L]$ grid, where $L =
(n'-2\alpha)\beta^{d-1}$. Actually it is not hard to check that
for the proof of the lower bound purpose, we can just think of the
new path set as being defined in the $[\alpha]^{d-1}\times [L]$
grid as in Section \ref{sec: weak grid}, with $V^w =
[\alpha]^{d-1}$ and $V^c = [L]$. \footnote{See Appendix C for more
explanations.} So we have $w_X = w_Y = \Omega(T)$, where $T = L-1
= \Theta(n^{1+(d-1)(1-r)})$. Also it holds $u_{X,x_{j,b}} \leq
O(\sum_{k=0}^{j+b-1} a_{k,j,b})$ and $v_{Y,x_{j,b}} =
O\left(\sum_{k=0}^{j-(d-1)\alpha^{2}+b-1}\frac{b_{k,j,b}}{\alpha^{d-1}}
+\sum_{k=j-(d-1)\alpha^{2}+b}^{j+b-2}\frac{b_{k,j,b}}{(j-k+b-1)^{-(d-1)/2}}+b_{j+b-1,j,b}\right)$.

For randomized lower bound, $a_{k,j,b} = b_{k,j,b} = 1$,
$v_{Y,x_{j,b}} = O(T/\alpha^{d-1} + \sum_{t=1}^{(d-1)\alpha^{2}}
t^{-(d-1)/2})$. So $w_Y/v_{Y,x_{j,b}} =
\Omega\left(\min\left\{n^{(d-1)r},
\frac{n^{1+(d-1)(1-r)}}{\sum_{t=1}^{(d-1)\lfloor n^{r}\rfloor ^2}
t^{-(d-1)/2}}\right\}\right)$ by noting that $\alpha =
\Theta(n^r)$ and $\beta = \Theta(n^{1-r})$. If $d\geq 4$, then let
$r = d/(2d-2)$ and we get $RLS([n]^d) = \Omega(N^{1/2})$. If
$d=3$, let $r=3/4 - \log\log n / (4\log n)$, and we get
$RLS([n]^3) = \Omega(N^{1/2}/\sqrt{\log N})$. For $d=2$, let $r =
2/3$ and we get $RLS([n]^2) = \Omega(N^{1/3})$.

For the quantum lower bounds, $u,v$ are defined as in \eqref{eq: def
u} and \eqref{eq: def v}, where $a_{k,j,b}$ is equal to $1$ if
$j+b-k=1$, equal to $(j-k+b-1)^{-(d-1)/4}$ if $1<j-k+b\leq
(d-1)\alpha^{2}$, and equal to $\alpha^{-(d-1)/2}$ if
$j-k+b>(d-1)\alpha^{2}$, and $b_{k,j,b} = a_{k,j,b}^{-1}$. Then
$QLS([n]^d) = \Omega\left(\min\left\{n^{(d-1)r/2},
\frac{n^{1+(d-1)(1-r)}}{\sum_{t=1}^{(d-1)\lfloor n^{r}\rfloor^2}
t^{-\frac{d-1}{4}}}\right\}\right)$. Now if $d\geq 6$, then letting
$r = 2d/(3d-3)$ and we get $QLS([n]^d) = N^{1/3}$. If $d=5$, then
let $r = 5/6-\log \log n / (6 \log n)$ and $QLS([n]^5) = (N/\log
N)^{1/3}$. For $2\leq d\leq 5$, we let $r = d/(d+1)$, then
$QLS([n]^d) = N^{1/2-1/(d+1)}$. This completes the proof of Theorem
\ref{thm: grid lb}.

\section{The new upper bound on the 2-dimensional grid}\label{sec:
upper bound}
In \cite{Aa04}, a quantum algorithm for Local Search
on general graphs is given as follows. Do a random sampling over
all the vertices, find the minimum $f$-value vertex $v$ in them
using the algorithm by Durr and Hoyer \cite{DH96} based on Grover
search \cite{Gr96}. If $v$ is a local minimal vertex, then return
$v$; otherwise we follow a \emph{decreasing path} as follows. Find
a neighbor of $v$ with the smallest $f$-value, and continue this
minimum-value-neighbor search process until getting to a local
minimum vertex. We can see that this algorithm is actually along
the generic algorithm approach (see Section 1), where the initial
point is picked as the best one over some random samples. Here our
idea is that after finding the minimum vertex of the sampled
points, in stead of following the decreasing path of it, we start
over within a smaller grid and do this recursively.

Now we describe the algorithm, with some notations as follows. For
a given function $f:[n]^2\rightarrow \mathbb{N}$, a vertex $v\in
[n]^2$ and a set $S\subseteq [n]^2$, let $n(v,S) = |\{u\in S:
f(u)<f(v)\}|$. A decreasing path of $v\in [n]^2$ is a sequence of
vertices $v_0=v, v_1,...,v_k$ such that $f(v_{i+1}) = \min_{v:
(v_i,v)\in E}f(v) < f(v_{i})$ for $i=0,...,k-1$ and $v_k$ is a
locally minimal vertex. We write $f(u)\leq f(S)$ if $f(u)\leq
f(v)$ for all $v\in S$. In particular, $f(u) \leq f(\emptyset)$ is
always true. For two vertices $u=(u_1,u_2),v=(v_1,v_2)\in [n]^2$,
the $l_1$ distance is $|u-v|_1 = |u_1-v_1| + |u_2-v_2|$.
In the following algorithm, the asymptotical numbers at the end of
some command lines are the numbers of quantum queries needed for
the line. For those commands without any number, no query is
needed.

\begin{enumerate}


\item $m_{(0)} = n$, $U_{(0)} = [n]^2$;

\item $i = 0$;

\item \textbf{while} ($|m_{(i)}| > \sqrt{n}$) \textbf{do}

  \begin{enumerate}
  \item \label{step: sample} Randomly pick (with replacement) $\lceil\frac{4|U_{(i)}|}{m_{(i)}}\log
  \frac{1}{\epsilon_1}\rceil$
  vertices from $U_{(i)}$;

  \item \label{step: min search} Search the sampled vertices for one $v_{(i)}$ with the minimal
  $f$ value, using Durr and Hoyer's algorithm \cite{DH96} with the success probability at least $1-\epsilon_2$.
  \qquad --- $O(\sqrt{\frac{4|U_{(i)}|}{m_{(i)}} \log \frac{1}{\epsilon_1}}\log \frac{1}{\epsilon_2})$

  \item \label{step: point update} \textbf{if} $i=0$, \textbf{then} $u_{(i+1)} = v_{(i)}$;

  \textbf{else} \textbf{if} $f(u_{(i)}) < f(v_{(i)})$, \textbf{then} $u_{(i+1)} =
  u_{(i)}$;

  \qquad \textbf{else} $u_{(i+1)} = v_{(i)}$;

  \item \label{step: boundary search} \textbf{for} $j = 1$ \textbf{to} $\log \frac{1}{\epsilon_3}$
  \begin{enumerate}
  \item \label{step: boundary sample} Randomly pick $m\in
  [\lfloor m_{(i)}/4\rfloor,\ \lceil3m_{(i)}/4\rceil]$, and let $W_{(i)} = \{w\in U_{(i)}: |w -
  u_{(i+1)}|_1 = m\}$.

  \item \label{step: boundary test} Test whether $f(u_{(i+1)}) \leq f(W_{(i)})$ by Grover Search over
  $W_{(i)}$ with the error probability less than $\epsilon_4$. \qquad --- $O(\sqrt{m_{(i)}}\log \frac{1}{\epsilon_4})$

  \item If the answer is Yes, jump out of this \textbf{for} loop and go to Step \ref{step: length and range update}.

  \end{enumerate}

  \item Report Fail and terminate the whole algorithm.

  \item \label{step: length and range update} $m_{(i+1)} = m$,
  $U_{(i+1)}= \{u\in U_{(i)}: |u-u_{(i+1)}|_1\leq m_{(i+1)}\}$;

  \item $i = i + 1$;
  \end{enumerate}

\item \label{step: final search} Follow a decreasing path of
$u_{(i)}$ to get a locally minimum vertex. \qquad ---
$O(\sqrt{n})$

\end{enumerate}

We now analyze the algorithm. Define the boundary $B(S)$ of a set
$S\subseteq [n]^2$ to be the set $\{s\in S: \exists t\in [n]^2-S,\
\st\ |s-t|_1 = 1\}$. Therefore, $B([n]^2) = \emptyset$.

Step \ref{step: sample} - \ref{step: point update}: Denote by $S$
the set of the $\lceil \frac{4|U_{(i)}|}{m_{(i)}}\log
\frac{1}{\epsilon_1}\rceil $ sampled vertices in Step \ref{step:
sample}, and let $a = \min_{u\in S} f(u)$. Then $|\{v\in U_{(i)}:
f(v) < a\}| \leq m_{(i)}/4$ with probability at least
$1-\epsilon_1$. Step \ref{step: min search} can find a $u$
achieving the minimum in the definition of $a$ with probability at
least $1-\epsilon_2$. Put the two things together, we have
$n(v_{(i)},U_{(i)})\leq m_{(i)}/4$ with probability at least
$1-\epsilon_1-\epsilon_2$. Since $f(u_{(i+1)}) \leq f(v_{(i)})$,
we have $n(u_{(i+1)},U_{(i)})\leq m_{(i)}/4$ with probability at
least $1-\epsilon_1-\epsilon_2$ too.

Step \ref{step: boundary search} - \ref{step: length and range
update}: In the event $n(u_{(i+1)},U_{(i)})\leq m_{(i)}/4$, at
most $m_{(i)}/4$ possible $m$'s in $[m_{(i)}]$ have ``$\exists
w\in W_{(i)}$ \st $f(w) < f(u_{(i+1)})$", because $\forall m\in
[m_{(i)}], W_{(i)}\subseteq U_{(i)}$, and different $W_{(i)}$'s
(due to different $m$) do not intersect. We say an $m$ is
\emph{good} (and also $W_{(i)}$ is \emph{good}) if $f(u_{(i+1)})
\leq f(W_{(i)})$. Therefore, at least $m_{(i)}/4$ number of $m$'s
in $[\lfloor m_{(i)}/4\rfloor ,\ \lceil 3m_{(i)}/4\rceil ]$ are
good. Since we pick $m$ for $\log \frac{1}{\epsilon_3}$ times, we
will get a good $m$ with probability $1-\epsilon_3$. The
probability that all the Grover searches in Step \ref{step:
boundary test} in the $\log \frac{1}{\epsilon_3}$ tries are
correct is at least $1-\epsilon_4 \log \frac{1}{\epsilon_3}$.
Putting the two things together, we know that if
$n(u_{(i+1)},U_{(i)})\leq m_{(i)}/4$, then with probability at
least $1-\epsilon_3 - \epsilon_4 \log \frac{1}{\epsilon_3}$, we
can reach Step \ref{step: length and range update} with a good $m$
found.

If both $n(u_{(i+1)},U_{(i)})\leq m_{(i)}/4$ happens and we find a
good $m$, then we have two consequences. The first one is that
$n(u_{(i+1)}, U_{(i+1)}) \leq m_{(i+1)}$. Actually, since
$U_{(i+1)} \subseteq U_{(i)}$, we have $n(u_{(i+1)},U_{(i+1)})\leq
\lfloor m_{(i)}/4 \rfloor \leq m_{(i+1)}$. (Since
$n(u_{(i+1)},U_{(i)})$ is an integer, $n(u_{(i+1)},U_{(i)})\leq
m_{(i)}/4$ is equivalent to $n(u_{(i+1)},U_{(i)})\leq \lfloor
m_{(i)}/4\rfloor $.) The second consequence is that $f(u_{(i+1)})
\leq f(B(U_{(i+1)}))$, provided that all the $W_{(0)}, ...,
W_{(i-1)}$ are good. To see this, we first show $B(U_{(i+1)})
\subseteq B(U_{(i)}) \cup W_{(i)}$. In fact, any $s\in
B(U_{(i+1)})$ satisfies that $s\in U_{(i+1)}$ and that $\exists
t\in [n]^2-U_{(i+1)}$ \st $|s-t|_1 = 1$. Recall that
$U_{(i+1)}\subseteq U_{(i)}$, so if $t\in [n]^2-U_{(i)}$, then
$s\in B(U_{(i)})$ by definition. Otherwise $t\in
U_{(i)}-U_{(i+1)}$, and thus $t\in U_{(i)}$ and $|t-u_{(i+1)}|_1 >
m_{(i+1)}$ by the definition of $U_{(i+1)}$. Noting that
$|s-u_{(i+1)}|_1 \leq m_{(i+1)}$ since $s\in U_{(i+1)}$, and that
$|s-t|_1 = 1$, we have $|s-u_{(i+1)}|_1 = m_{(i+1)}$, which means
$s\in W_{(i)}$. Thus for all $s\in B(U_{(i+1)})$, either $s\in
B(U{(i)})$ or $s\in W_{(i)}$ holds, which implies $B(U_{(i+1)})
\subseteq B(U{(i)})\cup W_{(i)}$. Continuing this process, we have
$B(U_{(i+1)}) \subseteq B(U_{(0)}) \cup W_{(0)} \cup...\cup
W_{(i)} = W_{(0)} \cup...\cup W_{(i)}$. Now to prove $f(u_{(i+1)})
\leq f(B(U_{(i+1)}))$, it is enough to show that $f(u_{(i+1)})
\leq f(W_1\cup...\cup W_{(i)})$. But this is easy by noting that
$f(u_{(i+1)}) \leq ... \leq f(u_{(1)})$, and that $f(u_{(j+1)})
\leq f(W_{(j)})$ for any $j=0,...,i$ because we assume that all
the $W_{(j)}$'s are good.

Putting all these together, we know that if all $W_{(0)}, ...,
W_{(i-1)}$ are good, then with probability
$1-(\epsilon_1+\epsilon_2+\epsilon_3+\epsilon_4\log
\frac{1}{\epsilon_3})$, we have that $W_{(i)}$ is good,
$n(u_{(i+1)},U_{(i+1)})\leq m_{(i+1)}$ and $f(u_{(i+1)}) \leq
f(B(U_{(i+1)}))$. Denote by $I$ the final value of $i$ (when the
algorithm jumps out of the \textbf{while} loop), and let $\epsilon
= \epsilon_1+\epsilon_2+\epsilon_3+\epsilon_4\log
\frac{1}{\epsilon_3}$. Then by a simple induction, we know that
with probability at least $1-I\epsilon$, we have that 1) all
$W_{(0)}, ..., W_{(I-1)}$ are good, 2) $n(u_{(I)},U_{(I)})\leq
m_{(I)}$ and 3) $f(u_{(I)}) \leq f(B(U_{(I)}))$. Note that 3)
implies that any decreasing path of $u_{(I)}$ cannot go out of the
$U_{(I)}$. Together with 2), we have that any decreasing path of
$u_{(I)}$ is no longer than $m_{(I)} \leq \sqrt{n}$, and thus
following the path will get to a locally minimum vertex by no more
than $\sqrt{n}$ queries.

Since $m_{(i+1)} \leq \lceil 3m_{(i)}/4\rceil$ (and $m_{(i)}
> \sqrt{n}$), we have $I \leq \log n$. Let $\epsilon =
\frac{1}{2\log n}$, $\epsilon_1 = \epsilon_2 = \epsilon_3 =
\epsilon/4 = O(1/\log n)$ and $\epsilon_4 =
\epsilon/(4\log\frac{4}{\epsilon}) = O(1/\log n \log\log n)$. Then
the algorithm can finds a locally minimum vertex with the
probability at least $1-I\epsilon = 1/2$.

As to the number of queries used, Step \ref{step: min search} uses
$O(\sqrt{\frac{4|U_{(i)}|}{m_{(i)}} \log \log n}\log \log n) =
O(\sqrt{m_{(i)}} (\log \log n)^{1.5})$ queries, because $|U_{(i)}|
\leq 2m_{(i)}^2$ (see Step \ref{step: length and range update}).
Step \ref{step: boundary search} uses $O(\log \log
n\sqrt{m_{(i)}}\log(\log n \log \log n)) =
O(\sqrt{m_{(i)}}(\log\log n)^2)$ queries. Finally, Step \ref{step:
final search} uses $O(\sqrt{n})$ queries. Altogether, the total
number of queries used is less than $O(\sum_{i=1}^{\log n}
\sqrt{m_{(i)}}(\log\log n)^{1.5} + \sqrt{m_{(i)}} (\log\log n)^2 +
\sqrt{n})= O(\sqrt{n}(\log\log n)^2)$.

\section{Concluding Remarks: further improvements and generalizations}\label{sec: conclusion}
The paper gives new lower bounds for Local Search problems. Some
other random walk can be used to further improve the lower bound
on low dimension grid cases. For example, by cutting the
2-dimensional grid into $n^{2/5}$ blocks (each of size
$n^{4/5}\times n^{4/5}$) and using a random walk similar to
Aaronson's in \cite{Aa04} (but with some modifications to make the
path self-avoiding), we can prove $QLS([n]^2) = N^{1/5}/\log N$.
But this walk suffers from the fact that the ``passing
probability" is now $n^{4/5}$ times the ``stopping probability".
So it only works better at dimension 2. We put the further results
in a complete version of the paper.

The lower bound technique we use can be easily generalized to the
Local Search on product graphs. Precisely, $G=(V,E)$ is a product
graph if $G$ can be decomposed as $G_1\times G_2 = \left(V_1\times
V_2, (E_1\times I_2) \cup (I_1\times E_2)\right)$ where $I_i =
\{(v_i,v_i):v_i\in V_i\}$ for $i=1,2$. Some graphs, like
hypercubes, may have many ways of decomposition. For a fixed
decomposition $\mathcal{D}$, suppose we have a random walk $W$ on
graph $G_1 = (V_1, E_1)$ with transition probability $\{p_{ij}: i
= j \text{ or } (i,j)\in E\}$ and stationary distribution $\pi$.
Denote by $p_{ij}^{(t)}$ the probability that the random walk
starting at $i$ stops at $j$ after $t$ steps. Let $p^{(t)} =
\max_{ij}p_{ij}^{(t)}$ and $\pi_{max} = \max_{i\in V_1}\pi(i)$. We
say the walk mixes at time $t_0$ if $p^{(t_0)} \leq 2\pi_{max}$.
Let $p_1 = \sum_{t\leq t_0} p^{(t)}$ and $p_{1/2} = \sum_{t\leq
t_0} \sqrt{p^{(t)}}$. Then under mild conditions, we have
\begin{equation}\label{eq: general graph lb}
RLS(G) = \Omega(\max_{\mathcal{D}}\min\{\frac{1}{\pi_{max}},
\frac{L}{p_1}\}), \qquad QLS(G) =
\Omega(\max_{\mathcal{D}}\min\{\frac{1}{\sqrt{\pi_{max}}},
\frac{L}{p_{1/2}}\})
\end{equation}
where $L$ is the length of the longest self-avoiding path in $G_2$.

Random walk has been widely studied as a sampling method for
algorithms, where the key parameter is the mixing time. It is
interesting that both Aaronson's \cite{Aa04} and this paper use
random walk to give lower bounds. And we can see from \eqref{eq:
general graph lb} that for lower bounds, we care not only about the
mixing time of the random walk, but also about its behavior before
mixing.

The paper also gives a quantum upper bound on 2-dimensional grid
Local Search. The technique naturally applies to the graph $G$
that expands slowly: if for any vertex $v$ and integer $k$, the
number of vertices that $v$ can reach within (and exactly by,
resp.) $k$ steps is at most $c(k)$, (and $r(k)$, resp), then
\begin{align}
RLS(G) = O\left(\sum_{i=0}^{\log d}
\left(\frac{c(m_{(i)})}{m_{(i)}} +
r(m_{(i)})\right)(\log\log d)^2\right), \qquad \\
\label{eq: quantum upper bound} QLS(G) = O\left(\sum_{i=0}^{\log
d} \left(\sqrt{\frac{c(m_{(i)})}{m_{(i)}}} +
\sqrt{r(m_{(i)})}\right)(\log\log d)^2 \right),
\end{align}
where $m_{(0)} = d$, the diameter of the graph, and $m_{(i)} \in
[m_{(i-1)}/4,\ 3m_{(i-1)}/4]$. For 2-dimensional grid, we have
$c(k) = \Theta(k^2)$ and $r(k) = \Theta(k)$, so \eqref{eq: quantum
upper bound} gives the upper bound in Section \ref{sec: upper
bound}.


\vspace{1em}\noindent\textbf{Acknowledgement}

The author thanks Scott Aaronson, Xiaoming Sun and Andy Yao very
much for many valuable discussions. Thanks also to Yves Verhoeven
for pointing out an error in the upper bound section in a previous
version.


\appendix \vspace{2em} \noindent
\begin{LARGE}\textbf{Appendix}\end{LARGE}

\section{Proof of Lemma \ref{lem: hypercube hit prob}}
Recall that suppose the $j$-th ball is put into $i_j$-th bin, and
$n_i \equiv 1$ means $n_i$ is odd, $n_i\equiv 0$ means $n_i$ is
even.

\begin{proof}
First, it is easy to see that for any $b_1,...,b_m\in \B$ and any
$i^*\in [m]$, it holds that
\begin{equation}
p_{i^*}^{(t)}[b_1,...,b_m] \leq \frac{m}{m-1} p^{(t)}[b_1,...,b_m]
\end{equation}
Actually,
\begin{align}
p^{(t)}[b_1,...,b_m] & = \pr[i_1 = i^*]\pr[(i_1,...,i_t) \text{
generates }(b_1,...,b_m)|i_1 = i^*] \\
& \quad + \pr[i_1 \neq i^*]\pr[(i_1,...,i_t) \text{
generates }(b_1,...,b_m)|i_1 \neq i^*]\\
& = \frac{1}{m}
p^{(t-1)}[b_1,...b_{i^*-1},1-b_{i^*},b_{i^*+1},...,b_m] +
\frac{m-1}{m}p_{i^*}^{(t)}[b_1,...,b_m]\\
& \geq \frac{m-1}{m}p_{i^*}^{(t)}[b_1,...,b_m]
\end{align}
So to prove the lemma, it is enough to show the same upper bound for
$p^{(t)}[b_1,...,b_m]$.

We start with several simple observations. First, we assume that $t$
and $\sum_{i=1}^m b_i$ have the same parity, because otherwise the
probability is 0 and the lemma holds trivially. Second, by the
symmetry, any permutation of $b_1,...,b_m$ does not change
$p^{(t)}[(b_1,...,b_m)]$. Third, $p^{(t)}[(b_1,...,b_m)]$ decreases
if we replace two 1's in $b_1,...,b_m$ by two $0$'s. Precisely, if
we have two $b_i$'s being 1, say $b_1 = b_2 = 1$, then
$[(b_1,...,b_m)] < p^{(t)}[(0,0,b_3,...,b_m)]$. In fact, note that
\begin{align}
p^{(t)}[(b_1,...,b_m)] & = \frac{1}{m^t}\sum_{\scriptstyle
n_1+...+n_m = t \atop n_i \equiv b_i, i\in
[m]}\frac{t!}{n_1!...n_m!} \\
& = \frac{1}{m^t}\sum_{\scriptstyle n_3+...+n_m\leq t  \atop
n_i\equiv b_i, i=3,...,m}\left(\frac{t!}{(n_1+n_2)!n_3!...n_m!}
\sum_{\scriptstyle n_1+n_2 = t-n_3-...-n_m \atop  n_i\equiv b_i, i =
1,2}\frac{(n_1+n_2)!}{n_1!n_2!}\right)
\end{align}
where as usual, let $0!=1$. If $n_3+...+n_m < t$, then
\begin{equation}
\sum_{\scriptstyle n_1+n_2 = t-n_3-...-n_m\atop n_i\equiv 1, i =
1,2}\frac{(n_1+n_2)!}{n_1!n_2!} = \sum_{\scriptstyle n_1+n_2 =
t-n_3-...-n_m\atop n_i\equiv 0, i = 1,2}\frac{(n_1+n_2)!}{n_1!n_2!}
\end{equation}
If $n_3+...+n_m = t$, then the only possible $(n_1,n_2)$ is $(0,0)$,
so
\begin{equation}
\sum_{\scriptstyle n_1+n_2 = t-n_3-...-n_m \atop n_i\equiv 1, i =
1,2}\frac{(n_1+n_2)!}{n_1!n_2!} = 0, \qquad \sum_{\scriptstyle
n_1+n_2 = t-n_3-...-n_m\atop n_i\equiv 0, i =
1,2}\frac{(n_1+n_2)!}{n_1!n_2!} = 1.
\end{equation}
Thus $p^{(t)}[(1,1,b_3,...,b_m)]< p^{(t)}[(0,0,b_3,...,b_m)]$.

By the observations, it is sufficient to prove the lemma for the
case $p^{(t)}[(0,...,0)]$ if $t$ is even, and for the case
$p^{(t)}[(1,0,...,0)]$ if $t$ is odd. Note that if $t$ is even, then
\begin{equation}
p^{(t)}[(0,...,0)] = \sum_{i=1}^m\pr[i_1 = i]\pr[(i_2,...,i_t)
\text{ generates } (e_i)]
\end{equation}
where $e_i$ is the $m$-long vector with only coordinate $i$ being 1
and all other coordinates being 0.  By the symmetry, $p^{(t-1)}[e_1]
= ... = p^{(t-1)}[e_{m}]$, thus $p^{(t)}[(0,...0)] = p^{(t-1)}[e_1]
= p^{(t-1)}[1,0,...,0]$. Therefore, it is enough to show the lemma
for even $t$.

We now express $p^{(t)}[0,...,0]$ in two ways. One is to prove the
first case ($t\leq 10$) in the lemma, and the other is for the
second case ($10<t\leq m^2$) and the third case ($t>m^2$) in the
lemma.

To avoid confusion, we write the number $m$ of bins explicitly as
subscript: $p_m^{(t)}[b_1,...,b_m]$. We consider which bin(s) the
first two balls is put into.
\begin{align}
p_m^{(t)}[0,...,0] & = \pr[i_1 = i_2]p_m^{(t-2)}[0,...,0] +
\pr[i_1\neq i_2] p_m^{(t-2)}[1,1,0,...,0]\\
& = \frac{1}{m}p_m^{(t-2)}[0,...,0] +
\frac{m-1}{m}p_m^{(t-2)}[1,1,0,...,0]
\end{align}
To compute $p_m^{(t-2)}[1,1,0,...,0]$, we consider to put $(t-2)$
balls in $m$ bins. By the analysis of the third observations above,
we know that
\begin{align}
& p_m^{(t-2)}[0,...,0] - p_m^{(t-2)}[1,1,0,...,0] \\
= & \pr[n_1 = n_2 = 0, n_3 \equiv 0, ..., n_m \equiv
0] \\
= & \pr[n_1 = n_2 = 0] \pr[n_3 \equiv 0, ..., n_m \equiv 0|n_1=n_2=0] \\
= & \left(\frac{m-2}{m}\right)^{t-2}p_{m-2}^{(t-2)}[0,...,0]
\end{align}
Therefore,
\begin{equation}
p_m^{(t)}[0,...,0] = p_m^{(t-2)}[0,...,0] -
\frac{m-1}{m}\left(\frac{m-2}{m}\right)^{t-2}p_{m-2}^{(t-2)}[0,...,0]
\end{equation}

Now using the above recursive formula and the base case
$p_m^{(2)}[0,...,0] = 1/m$, it is easy (but tedious) to prove by
calculations that $p_m^{(t)}[0,...,0] =
((t-1)!!/m^{\frac{t}{2}})(1-o(1))$ for even $t\leq 10$. This proves
the first case in the lemma.

For the rest two cases, consider the generating function
$(x_1+...+x_m)^t = \sum_{n_1+...+n_m =
t}\binom{t}{n_1,...,n_m}x_1^{n_1}...x_m^{n_m}$. If $x_i\in
\{-1,1\}$, then $(x_1+...+x_m)^t = \sum_{n_1+...+n_m =
t}\binom{t}{n_1,...,n_m}(-1)^{|\{i: x_i = -1, n_i\equiv 1\}|}$. We
sum it over all $x_1...x_n\in \{-1,1\}^n$. Note that for those
$(n_1,...,n_m)$ that has some $n_{i_0}\equiv 1$, it holds due to the
cancelation that $\sum_{x_1,...,x_m\in \{-1,1\}} (-1)^{|\{i: x_i =
-1, n_i\equiv 1\}|} = 0$ . On the other hand, if all $n_i$'s are
even, then $\sum_{x_1,...,x_m\in \{-1,1\}} (-1)^{|\{i: x_i = -1,
n_i\equiv 1\}|} = 2^m$. Thus we have $\sum_{x_1,...,x_m\in
\{-1,1\}}(x_1+...+x_m)^t = 2^m\sum_{\scriptstyle n_1+...+n_m =
t\atop n_i\equiv 0, i\in [m]}\binom{t}{n_1,...,n_m}$. Therefore
\begin{align}
p^{(t)}[0,...,0] & = \frac{1}{m^t}\sum_{\scriptstyle n_1+...+n_m =
t\atop n_i\equiv 0, i\in [m]}\binom{t}{n_1,...,n_m}\\
& = \frac{1}{2^m m^t} \sum_{x_1,...,x_m\in
\{-1,1\}}(x_1+...+x_m)^t \\
& = \frac{1}{2^m m^t}\sum_{i=0}^m\binom{m}{i}(m-2i)^t = \frac{1}{2^m
}\sum_{i=0}^m\binom{m}{i}\left(1-\frac{2i}{m}\right)^t.
\end{align}
It follows that $p^{(t)}[0,...,0]$ decreases with $t$, and this
proves the second case of the lemma with the help of the first case.
And if $t>m^2/2$, then
\begin{equation}
p^{(t)}[0,...,0] \leq \frac{1}{2^m
}\left(2+\left(1-\frac{2}{m}\right)^t\sum_{i=1}^{m-1}\binom{m}{i}\right)
<
 2/2^m+e^{-m} = O(1/2^m)
\end{equation}
This proves the third case of the lemma.
\end{proof}

\section{Proof of Proposition \ref{prop: short walk mixing}}
\begin{proof}
We consider two settings. One is as in the definition of the short
walk, where we have only $n$ points $0, ..., n-1$, and points $0$
and $n-1$ are two barriers\footnote{Here we let the $n$ points be
$0,...,n-1$ instead of $1,...,n$ just to make the later calculation
cleaner}. Another is the same except that the barriers are removed,
and we have infinite points in a line. For each $t$-bit binary
string $x=x_1...x_t$, we use $P_i^x$ and $Q_i^x$ to denote the two
paths that starting at $i$ and walk according to $x$ in the two
settings. Precisely, at step $s$, $Q_i^x$ goes left if $x_s = 0$ and
goes right if $x_s = 1$ . $P_i^x$ goes in the same way except that
it will stand still if the point is currently at left (or right) end
and it still wants to go left (or right). If the end point of
$P_i^x$ is $j$, then we write $i\rightarrow_t^{P,x} j$. Let
$X_{ij}^{(t),P}$ be the set of $x\in\B^t$ \st $i\rightarrow_t^{P,x}
j$, and put $n_{ij}^{(t),P}=|X_{ij}^{(t),P}|$. Then by definition,
$p_{ij}^{(t)} = n_{ij}^{(t),P}/2^t$. The notations
$i\rightarrow_t^{Q,x} j$, $X_{ij}^{(t),Q}$ and $n_{ij}^{(t),Q}$ are
similarly defined, with the corresponding $P$ changed to $Q$. Note
that $n_{ij}^{(t),Q} = \binom{t}{t/2 + (j-i)/2}$ if $j-i$ and $t$
have the same parity, and 0 otherwise. We now want to upper bound
$n_{ij}^{(t),P}$ in terms of $n_{ij}^{(t),Q}$.

For a path $P_i^x$, if at some step it is at point $0$ and wants to
go left, we say it \emph{attempts to pass the left barrier}.
Similarly for the right barrier. We say a path is in the $\{a_s,
b_s\}_{s=1}^l$ category if it first attempts to pass the left
barrier for $a_1$ times, and then attempts to pass the right barrier
for $b_1$ times, and so on. We call each round a stage $s$, which
begins at the time that $P_i^x$ attempts to pass the left barrier
for the $(a_1+...+a_{s-1}+1)$-th time, and ends right before the
time that $P_i^x$ attempts to pass the left barrier for the
$(a_1+...+a_{s}+1)$-th time. We also split each stage $s$ into two
halves, cutting at the time right before the path attempts to pass
the right barrier for the $(b_1+...+b_{s-1}+1)$-th time. Note that
$a_1$ may be 0, which means that the path first attempts to pass the
right barrier. Also $b_l$ may be $0$, which means the the last
barrier the path attempts to pass is the left one. But all other
$a_i, b_i$'s are positive. Also note that in the case of $l=0$, the
path never attempts to pass either barrier. We partition
$X_{ij}^{(t),P}$ as
\begin{equation}
X_{ij}^{(t),P} = \bigcup_{l,\
\{a_s,b_s\}_{s=1}^l}X_{ij}^{(t),P}[\{a_s,b_s\}_{s=1}^l]
\end{equation}
where $X_{ij}^{(t),P}[\{a_s,b_s\}_{s=1}^l]$ contains those paths
in the category $\{a_s,b_s\}_{s=1}^l$. Put
$n_{ij}^{(t),P}[\{a_s,b_s\}_{s=1}^l] =
|X_{ij}^{(t),P}[\{a_s,b_s\}_{s=1}^l]|$, thus
$n_{ij}^{(t),P}=\sum_{l}\sum_{\{a_s,b_s\}_{s=1}^l}n_{ij}^{(t),P}
[\{a_s,b_s\}_{s=1}^l]$.

Now consider the corresponding paths in $X_{ij}^{(t),Q}$. The
following observation relates $P_i^x$ and $Q_i^x$.

\begin{Obs} For each $x\in X_{ij}^{(t),P}[\{a_s,b_s\}_{s=1}^l]$,
the following two properties hold.
\begin{enumerate}
\item In the first half of stage $s$, the path $Q_i^x$ touches
(from right) but does not cross the point $\alpha_s=
\sum_{r=1}^{s-1}(b_r-a_r)-a_s$.

\item In the second half of stage $s$, the path $Q_i^x$ touches
(from left) but does not cross the point $\beta_s =
n-1+\sum_{r=1}^{s}(b_r-a_r)$

\item The path $Q_i^x$ ends at $\gamma =
j+\sum_{s=1}^{l}(b_s-a_s)$
\end{enumerate}
\end{Obs}

We let $Y_{i\gamma}^{(t),Q}[\{\alpha_s,\beta_s\}_{s=1}^l]$ contain
those $x\in \B^t$ satisfying the three conditions in the above
observation, and denote by
$m_{i\gamma}^{(t),Q}[\{\alpha_s,\beta_s\}_{s=1}^l]$ the size of the
set $Y_{i\gamma}^{(t),Q}[\{\alpha_s,\beta_s\}_{s=1}^l]$. Thus the
observation says
$X_{ij}^{(t),P}[\{\alpha_s,\beta_s\}_{s=1}^l]\subseteq
Y_{ij}^{(t),Q}[\{\alpha_s,\beta_s\}_{s=1}^l]$, and therefore we have
$n_{ij}^{(t),P}[\{a_s,b_s\}_{s=1}^l]\leq
m_{i\gamma}^{(t),Q}[\{\alpha_s,\beta_s\}_{s=1}^l]$. Now for each
$x\in Y_{i\gamma}^{(t),Q}[\{\alpha_s,\beta_s\}_{s=1}^l]$, if we
change the condition 1 in the case $s=1$ by allowing the path to
cross the point $\alpha_1$, and let
$Z_{i\gamma}^{(t),Q}[\{\alpha_s,\beta_s\}_{s=1}^l]$ be the new set
satisfying the new conditions, then
$m_{i\gamma}^{(t),Q}[\{\alpha_s,\beta_s\}_{s=1}^l] =
|Z_{i\gamma}^{(t),Q}[\{\alpha_s,\beta_s\}_{s=1}^l]| -
|Z_{i\gamma}^{(t),Q}[\alpha_1-1, \beta_1,
\{\alpha_s,\beta_s\}_{s=2}^l]|$. In other words, the set of paths
touches (from right) but does not cross $\alpha_1$ is the set of
paths touches or crosses $\alpha_1$ minus the set of paths touches
or crosses $\alpha_1-1$.

Now we calculate
$|Z_{i\gamma}^{(t),Q}[\{\alpha_s,\beta_s\}_{s=1}^l]|$ by the
so-called reflection rule. Suppose the first time that $Q_i^x$
touches $\alpha_1$ is $t_{1}$. We reflect the first $t_1$ part of
the path $Q_i^x$ with respect to the point $\alpha_1$. Precisely,
let $y = (1-x_1)...(1-x_{t_{1}})x_{t_{1}+1}...x_t$, then the paths
$Q_i^x$ and $Q_{2\alpha_1-i}^y$ merge at time $t_{1}$. And it is
easy to check that it is a 1-1 correspondence between
$Z_{i\gamma}^{(t),Q}[\{\alpha_s,\beta_s\}_{s=1}^l]$ and
$Y_{2\alpha_1-i,\gamma}^{(t),Q}[\beta_1,\{\alpha_s,\beta_s\}_{s=2}^l]$,
Here
$Y_{2\alpha_1-i,\gamma}^{(t),Q}[\beta_1,\{\alpha_s,\beta_s\}_{s=2}^l]$
is the set of paths starting at $2\alpha_1-i$, satisfying (a) the
condition 2 at the first stage, (b) both conditions 1 and 2 at the
rest $l-1$ stages, and (c) condition 3. So
\begin{align}
|Z_{i\gamma}^{(t),Q}[\{\alpha_s,\beta_s\}_{s=1}^l]| & =
|Y_{2\alpha_1-i,\gamma}^{(t),Q}[\beta_1,\{\alpha_s,\beta_s\}_{s=2}^l]|
=
m_{2\alpha_1-i,\gamma}^{(t),Q}[\beta_1,\{\alpha_s,\beta_s\}_{s=2}^l]
\\ \label{eq: alpha}
& = m_{-2a_1-i,\gamma}^{(t),Q}[\beta_1,\{\alpha_s,\beta_s\}_{s=2}^l]
\\ \label{eq: move}
& =
m_{-a_1-i,\gamma+a_1}^{(t),Q}[\beta_1+a_1,\{\alpha_s+a_1,\beta_s+a_1\}_{s=2}^l]
\end{align}
where \eqref{eq: alpha} is due to the fact that $\alpha_1 = -a_1$,
and \eqref{eq: move} is because that the number of the paths does
not change if we move all the paths right by $a_1$. Similarly, we
have
\begin{align}
|Z_{i\gamma}^{(t),Q}[\alpha_1-1, \beta_1,
\{\alpha_s,\beta_s\}_{s=2}^l]| & =
m_{2\alpha_1-2-i,\gamma}^{(t),Q}[\beta_1,\{\alpha_s,\beta_s\}_{s=2}^l] \\
& =
m_{-a_1-2-i,\gamma+a_1}^{(t),Q}[\beta_1+a_1,\{\alpha_s+a_1,\beta_s+a_1\}_{s=2}^l]
\end{align}
Therefore,
\begin{align}
n_{ij}^{(t),P} [\{a_s,b_s\}_{s=1}^l] & \leq
m_{i\gamma}^{(t),Q}[\{\alpha_s,\beta_s\}_{s=1}^l] \\
& =
m_{-2a_1-i,\gamma}^{(t),Q}[\beta_1,\{\alpha_s,\beta_s\}_{s=2}^l] -
m_{-2a_1-2-i,\gamma}^{(t),Q}[\beta_1,\{\alpha_s,\beta_s\}_{s=2}^l]
\\
& =
m_{-a_1-i,\gamma+a_1}^{(t),Q}[\beta_1+a_1,\{\alpha_s+a_1,\beta_s+a_1\}_{s=2}^l]
\\ & \quad -
m_{-a_1-2-i,\gamma+a_1}^{(t),Q}[\beta_1+a_1,\{\alpha_s+a_1,\beta_s+a_1\}_{s=2}^l]
\end{align}
Now for any fixed $l>1$, we consider those categories with $a_1>0$
and $b_l>0$. Other cases can handled similarly. Note that
$\alpha_s+a_1 = b_1+\sum_{r=2}^{s-1} (b_r-a_r)-a_s$, \ $\beta_s+a_1
= n-1+\sum_{r=2}^s(b_r-a_r)$ and $\gamma+a_1 =
j+b_1+\sum_{r=2}^s(b_r-a_r)$ are all functions of $(b_1, a_2,b_2,
..., a_l, b_l)$, not of $a_1$ any more. Therefore,
\begin{align}
& \sum_{a_1,b_1,...,a_l,b_l>0} n_{ij}^{(t),P}[\{a_s,b_s\}_{s=1}^l] \\
\leq & \sum_{b_1,...,a_l,b_l>0}\sum_{a_1>0}
(m_{-a_1-i,\gamma+a_1}^{(t),Q}[\beta_1+a_1,\{\alpha_s+a_1,\beta_s+a_1\}_{s=2}^l]
\\ & \qquad \qquad \qquad \qquad -
m_{-a_1-2-i,\gamma+a_1}^{(t),Q}[\beta_1+a_1,\{\alpha_s+a_1,\beta_s+a_1\}_{s=2}^l])
\\
= & \sum_{b_1,...,a_l,b_l>0}
(m_{-1-i,\gamma+a_1}^{(t),Q}[\beta_1+a_1,\{\alpha_s+a_1,\beta_s+a_1\}_{s=2}^l]
\\
& \qquad \qquad \qquad
+m_{-2-i,\gamma+a_1}^{(t),Q}[\beta_1+a_1,\{\alpha_s+a_1,\beta_s+a_1\}_{s=2}^l])
\end{align}
Note that due to the parity, only one of
$m_{-1-i,\gamma+a_1}^{(t),Q}[\beta_1+a_1,\{\alpha_s+a_1,\beta_s+a_1\}_{s=2}^l]$
and
$m_{-2-i,\gamma+a_1}^{(t),Q}[\beta_1+a_1,\{\alpha_s+a_1,\beta_s+a_1\}_{s=2}^l]$
is nonzero. So the summation of them two items is equal to the
maximum of them. Now using the similar methods, \ie reflecting with
respect to points $(n-1+b_1)$ and $(n+b_1)$, moving the paths left
by $b_1$, and finally collapsing the telescope, we can get
\begin{align}
& \sum_{b_1,...,a_l,b_l>0}
m_{-1-i,\gamma+a_1}^{(t),Q}[\beta_1+a_1,\{\alpha_s+a_1,\beta_s+a_1\}_{s=2}^l]
\\
= & \sum_{a_2,b_2,...,a_l,b_l>0}
\max\{m_{2n+i,\gamma+a_1-b_1}^{(t),Q}[\{\alpha_s+a_1-b_1,\beta_s+a_1-b_1\}_{s=2}^l],\\
& \qquad \qquad \qquad \qquad
m_{2n+i+1,\gamma+a_1-b_1}^{(t),Q}[\{\alpha_s+a_1-b_1,\beta_s+a_1-b_1\}_{s=2}^l]\}
\end{align}
and
\begin{align}
& \sum_{b_1,...,a_l,b_l>0}
m_{-2-i,\gamma+a_1}^{(t),Q}[\beta_1+a_1,\{\alpha_s+a_1,\beta_s+a_1\}_{s=2}^l]
\\
= & \sum_{a_2,b_2,...,a_l,b_l>0}
\max\{m_{2n+i+2,\gamma+a_1-b_1}^{(t),Q}[\{\alpha_s+a_1-b_1,\beta_s+a_1-b_1\}_{s=2}^l],\\
& \qquad \qquad \qquad \qquad
m_{2n+i+3,\gamma+a_1-b_1}^{(t),Q}[\{\alpha_s+a_1-b_1,\beta_s+a_1-b_1\}_{s=2}^l]\}
\end{align}
We continue this process, and finally it is
\begin{align}
\sum_{a_1,b_1,...,a_l,b_l>0} n_{ij}^{(t),P}[\{a_s,b_s\}_{s=1}^l] &
\leq \max\{n_{2ln+i+h,\gamma+\sum_{s=1}^l(a_s-b_s)}^{(t),Q}:
h=0,...,4l-1\} \\
& = \max\{n_{2ln+i+h,j}^{(t),Q}: h=0,...,4l-1\}\\
& = n_{2ln+i,j}^{(t),Q} \\
& \leq \binom{t}{\frac{t}{2}+\frac{j-i-2ln}{2}}
\end{align}
Thus
\begin{equation}
\sum_{l>0}\sum_{a_1,b_1,...,a_l,b_l>0}
n_{ij}^{(t),P}[\{a_s,b_s\}_{s=1}^l] \leq \sum_{l>0}
\binom{t}{\frac{t}{2}+\frac{j-i}{2}-ln} =
\begin{cases}
O(\frac{2^t}{\sqrt{t}}) & \text{ if } t<n^2 \\
O(\frac{\sqrt{t}}{n}\frac{2^t}{\sqrt{t}}) = O(\frac{2^t}{n}) &
\text{ if }t\geq n^2
\end{cases}
\end{equation}

For other categories that $a_1 = 0$ or $b_l = 0$, the same result
can be proved similarly, and the $l=0$ is easy since $n_{ij}^{(t),Q}
= O(2^t/\sqrt{t})$. Putting all things together, we get the result
\begin{equation}
p_{ij}^{(t)} =
\begin{cases}
O(1/\sqrt{t}) & \text{if}\quad t\leq n^2 \\
O(1/n) & \text{if}\quad t > n^2 \\
\end{cases}
\end{equation}
for any $i$ and $j$, which completes our proof.
\end{proof}

\section{Further explanations of the construction in Section \ref{sec: better
grid}} In this section, we further explain the construction in
Section \ref{sec: better grid}. In particular, we shall make the
claim more clear that we can think of the construction the same as
a long grid $[\alpha]^{d-1}\times [(n'-2\alpha)\beta^{d-1}]$.
Actually, what we care about is, as before, the probability that
the random walk starting from a point $x = x_0...x_{d-1}$ passes
another point $x' = x_0...x_{d-1}$ after exactly $t'-t$ steps.
Here $t$ is the time that the random path passes $x$ and $t'$ is
the time that the path passes $x'$. Note that $t$ is fixed and
known by $x$ itself; similarly for $t'$. Denote this probability
by $\pr[x\rightarrow x']$. Suppose $x_i = (k_i-1)\alpha+y_i$ and
$x'_i = (k'_i-1)\alpha+y'_i$ for $i\in \{0,...,d-2\}$.

We first consider the case that one of the two points, say $x'$ is
on a block-changing segment. Since different block-changing
segments never intersect, a path passes $x'$ if and only if the
path passes the boundary point $x''$ at the beginning of the
block-changing segment that $x'$ is in. Also note that the time
that the path passes $x''$ is also $t'$ because the time does not
elapse on the block-changing segment. So it holds that
$\pr[x\rightarrow x'] = \pr[x\rightarrow x'']$, and it is enough
to consider the case that both $x$ and $x'$ are not in
clock-changing segments.

Now suppose both $x$ and $x'$ are not in clock-changing segments.
In general, $x$ and $x'$ may be not in the same block , so going
from $x$ to $x'$ needs to change blocks. Recall that to change
from the block $(k_0,...,k_{d-2})$ to the next one, only one $k_i$
changes by increasing or decreasing by 1. Suppose that to go to
$x'$ from $x$, we change blocks for $c$ times, by changing
$k_{i_1}, k_{i_2}, ..., k_{i_c}$ in turn. Let $n_j = |\{s\in [c]:
i_s = j\}|$. Note that to get to $x'$ from $x$ after $t'-t$ steps,
the coordinate $j$ needs to be $x'_j$ after $t'-t$ steps for each
coordinate $j\in \{0,...,d-2\}$. It is not hard to see that if a
block-changing needs to change $k_j$, then only the coordinate $j$
gets reflected within the current block. That is, suppose the
coordinate $j$ is $(k_j-1)\alpha + y_j$ before the block-changing,
then it changes to $(k_j-1)\alpha + \alpha+1-y_j$ after the
block-changing. So if $c=1$, then $\pr[x\rightarrow x']$ is equal
to the probability that a random walk in $[\alpha]^{d-1}$ starting
from $y_0...y_{d-2}$ stops at $y''_0...y''_{d-2}$ after $t'-t$
steps, where $y''_j = y'_j$ if $j\neq i_1$ and $y''_{i_1}
=(k_{i_1}-1)\alpha + \alpha+1-y'_{i_1}$. For general $c$,
$\pr[x\rightarrow x']$ is equal to the probability that a random
walk in $[\alpha]^{d-1}$ starting from $y_0...y_{d-2}$ stops at
$y''_0...y''_{d-2}$ after $t'-t$ steps, where $y''_j = y'_j$ if
$n_j$ is even and $y''_{j} =(k_{j}-1)\alpha + \alpha+1-y'_{j}$ if
$n_j$ is odd. Note that the latter probability has nothing to do
with the block-changing; it is just the same as we have a clock
space $[(n'-2\alpha)\beta^{d-1}]$ to record the random walk on
$[\alpha]^{d-1}$. Thus we can use Proposition \ref{prop: short
walk mixing} to upper bound this probability and further the proof
of the lower bound.

\end{document}